\documentclass[%
 reprint,
 amsmath,amssymb,
 aps,
pra,
]{revtex4-1}

\usepackage{graphicx}
\usepackage{dcolumn}
\usepackage{bm}
\usepackage[colorlinks=true,citecolor=blue,linkcolor=magenta]{hyperref}
\usepackage{enumitem} 
\usepackage{dcolumn}
\usepackage{bm}
\usepackage[bbgreekl]{mathbbol}
\usepackage{amsmath,dsfont}
\usepackage[normalem]{ulem}
\usepackage{color}

\newcommand{\appropto}{\mathrel{\vcenter{
  \offinterlineskip\halign{\hfil$##$\cr
    \propto\cr\noalign{\kern2pt}\sim\cr\noalign{\kern-2pt}}}}}

\usepackage{amsmath}
\usepackage{bm}
\begin{document}

\preprint{APS/123-QED}

\title{Robustness to spontaneous emission of a variational quantum algorithm}
\date{\today}
\author{Lo\"{i}c Henriet}
\affiliation{Pasqal, 2 avenue Augustin Fresnel, 91120 Palaiseau, France}
\begin{abstract}
We study theoretically the effects of dissipation on the performances of a variational quantum algorithm used to approximately solve a combinatorial optimization problem, the Maximum Independent Set, on a platform of neutral atoms. We take a realistic model of dissipation with spontaneous emission, and show numerically that the detrimental effects of these incoherent processes are spontaneously attenuated by the variational nature of the procedure. We furthermore implement an alternative optimization scheme recently proposed, where one uses the expected shortfall of the mean energy as the objective function.
\end{abstract}
\maketitle

\section{Introduction}

\begin{figure}[t!]
\center
\includegraphics[scale=0.4]{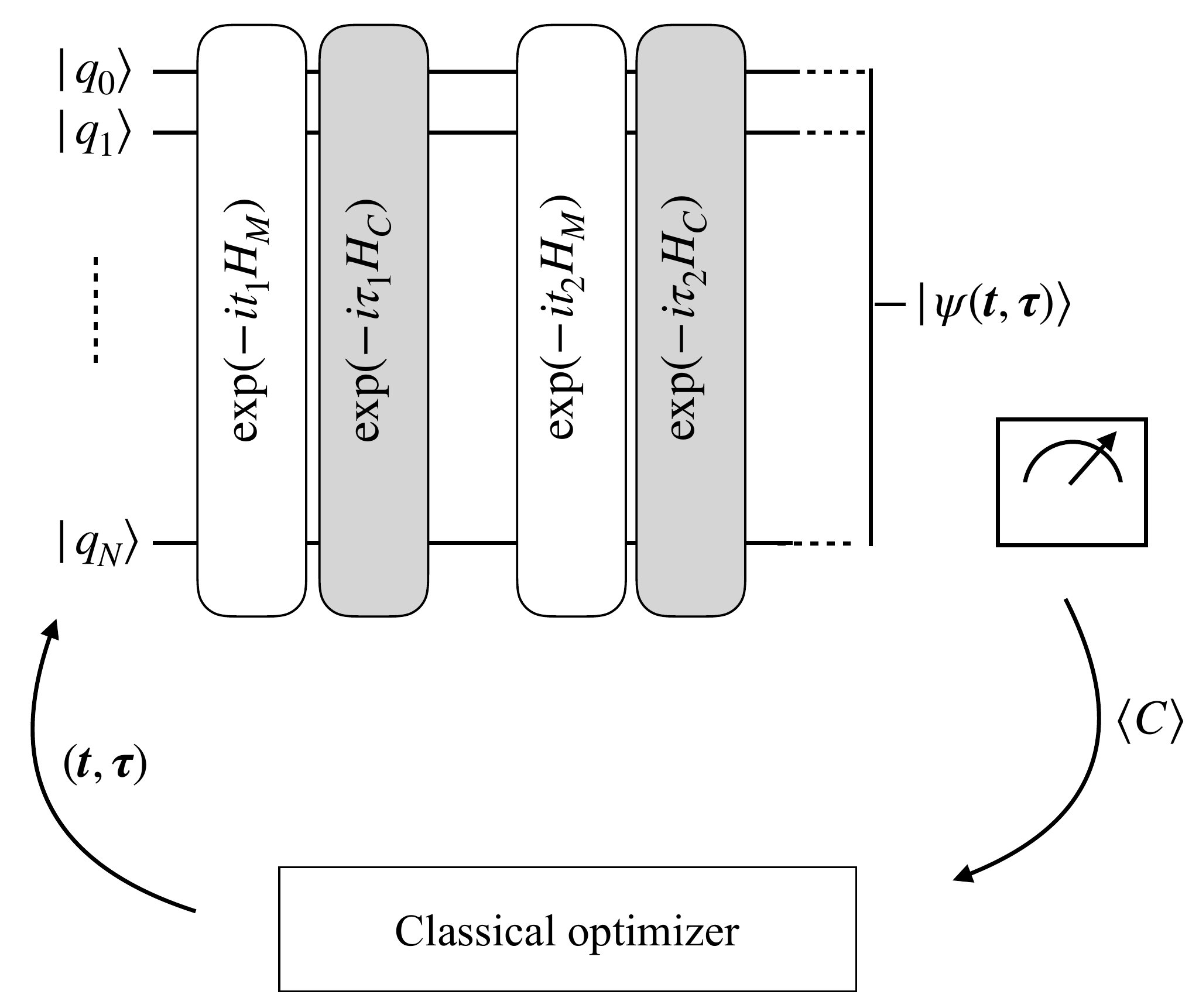}
\caption{Principle of hybrid quantum-classical variational algorithms. These algorithms are composed of both a quantum and a classical processor, that exchange information within a feedback loop. The quantum processor is used to prepare and measure a $N$-qubit parameterized wavefunction $|\psi(\bm{t},\bm{\tau})\rangle$, starting from the initial state $|q_0,q_1,...,q_N\rangle$. The outcome of the measurement $\langle C\rangle$ is then used as the objective function in a standard classical optimization procedure, that updates the parameters $\bm{t}$ and $\bm{\tau}$ for the next iteration. In this paper, the parameterized quantum circuit representing the action of the quantum processor is a succession of time evolutions with Hamiltonians $H_M$ and $H_C$.  }
\label{variational}
\end{figure}

Variational Quantum Computing (VQC) has recently emerged as a promising computing framework for noisy intermediate scale quantum processors (NISQ) \cite{Preskill2018}. The aim of such algorithms is to optimize for a given objective function $C$ resorting to both a quantum and a classical processor, within a closed feedback loop, and an example of such procedure is illustrated in Fig.\,\ref{variational}. In a first step, the quantum processor is harnessed to prepare and measure a trial wavefunction $|\psi(\bm{t},\bm{\tau})\rangle$, parameterized by $\bm{t}$ and $\bm{\tau}$. In this article, we consider that $|\psi(\bm{t},\bm{\tau})\rangle$ is the output of a parameterized quantum circuit, or quantum neural network, consisting of a sequence of time-evolutions with non-commuting Hamiltonians $H_M$ and $H_C$ during evolution times determined by $\bm{t}=(t_1,t_2,...,t_p)$ and $\bm{\tau}=(\tau_1,\tau_2,...,\tau_p)$, as illustrated in Fig.\,\ref{variational}. The dimension $p$ of $\bm{t}$ and $\bm{\tau}$ is referred to as the depth of the algorithm. From the outcome of the measurement, one computes an estimate $\langle C \rangle$ of the objective function to be optimized for. This estimate is then used as an input in a classical optimization procedure, which updates the variational parameters $\bm{t}$ and $\bm{\tau}$ for the following iteration. This loop is repeated multiple times until convergence is realized towards a final quantum state, from which one extracts an approximate solution of the problem under study.  In this framework, the quantum hardware determines the set of achievable states, while the classical optimization procedure controls whether the algorithm actually returns the best attainable solution within these achievable states.\\

Several versions of this idea have been developed for various applications, such as computing the ground state energy of a molecule\,\cite{Peruzzo2014,McClean_2016,Cao19}, solving non-linear partial differential equations\,\cite{Lubasch2019}, or finding the solution to combinatorial optimization problems\,\cite{Farhi2014}. Successful implementations of quantum chemistry calculations have been reported in photonic systems\,\cite{Peruzzo2014}, superconducting qubits platforms \,\cite{OMalley16,Kandala2017,Colless2018} and trapped ions \cite{Hempel2018}. A similar framework was used to compute the ground state of the lattice Schwinger model with trapped ions \,\cite{Kokail2019}.\\ 

It remains an open question whether such variational algorithms may display a quantum advantage as compared to standard classical algorithms\,\cite{Preskill2018,Guerreschi2019}. In particular, the possible quantum advantage of such algorithms for solving the combinatorial optimization problem MAX-3-LIN-2\,\cite{Farhi2014} at short depth was later negated by the discovery of efficient classical algorithms\,\cite{Boaz2015,Hastings19}. These variational algorithms triggered nonetheless a large interest, as they are believed to be resilient to errors, because of their approximate nature and the short depth of the circuits used, making them prime candidates for implementation on near-term quantum processors. However, little is known regarding the precise effects of errors and imperfections on these procedures. It was previously shown in Ref.\,\cite{OMalley16} that coherent errors or offsets did not affect overall performances. In addition, Ref.\,\cite{McClean2017} showed that the detrimental effects of incoherent errors could be suppressed by adding an additional step in the trial wavefunction preparation for quantum chemistry problems. It was also suggested recently to mitigate the effects of quantum noise by extrapolation to zero noise via Richardson deferred approach to the limit\,\cite{Temme2017}. \\

In this paper, we study an implementation of VQC with neutral atoms, as suggested in Refs.\,\cite{Pichler2018,Zhou2018}. More precisely, we will be interested in finding the solution to a particular combinatorial optimization problem, the Unit Disk Maximum Independent Set, under the presence of dissipation. We will show in particular that the algorithm under study is resilient to spontaneous emission since its variational nature attenuates the detrimental effects of noise. \\

The paper is structured as follows. In Section II, we introduce the Maximum Independent Set problem and the implementation of a VQC algorithm targeted to solve it with two-dimensional (2D) arrays of neutral atoms. In Section III, we study the effects of spontaneous emission on the performances of the algorithm and identify numerically a mechanism of self-mitigation of errors. In Section IV, we propose some changes to standard prescriptions in order to further improve the performance of the procedure. The code used for the numerical simulations of the variational algorithm is available online \,\cite{Code}.


\section{UD-MIS with Rydberg atoms}

In this Section, we introduce the Maximum Independent Set problem (MIS), and show how it can be approximately solved with a variational algorithm implemented on a quantum processor made of neutral atoms.\\

\begin{figure}[t!]
\center
\includegraphics[scale=0.5]{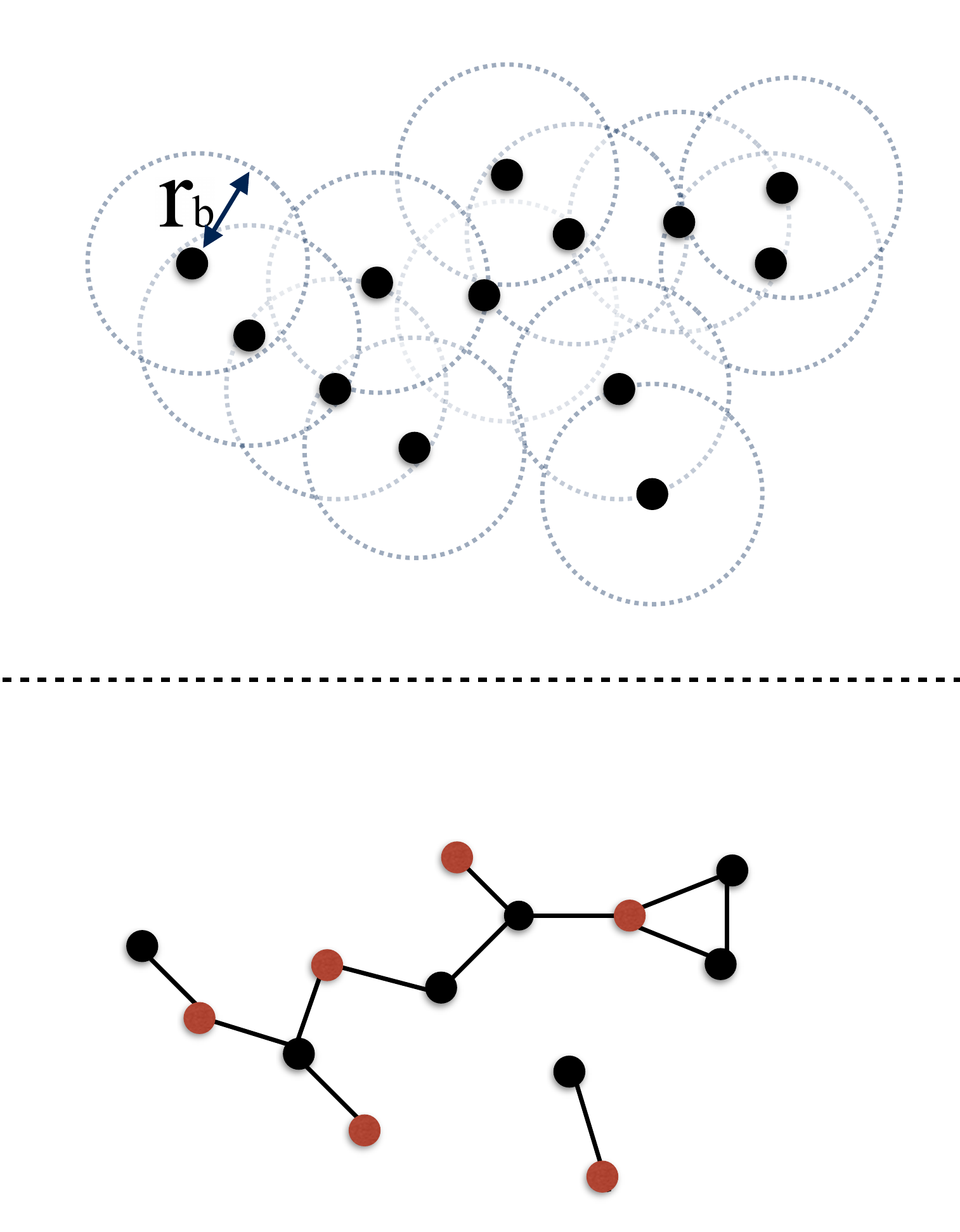}
\caption{The upper panel shows an ensemble of neutral atoms in 2D with random initial positions, together with their Rydberg blockade region characterized by the Rydberg blockade radius $r_b$ (see Appendix \ref{appendix_Rydberg}). It defines a graph in 2D where vertices are connected if they are closer than $r_b$, shown on the bottom panel. We display on that graph an example of an independent set in red. As there are no edges connecting two red vertices, this subset constitutes an independent set. The strong Rydberg-Rydberg interaction naturally prevents the excitation of connected atoms, so that the dynamics generated by $H$ in Eq. (\ref{ising}) is naturally restricted to independent sets.  }
\label{UD}
\end{figure}

Let us consider an undirected graph composed of a set of vertices connected by unweighted edges. An independent set of this graph is a subset of vertices where any two elements of this subset are not connected by an edge (see example of an independent set in Fig.~\ref{UD}, bottom panel). The objective of the MIS problem is to find the largest of such subsets. Deciding whether the size of the MIS is larger than a given integer is a NP-complete problem\,\cite{Garey79}. The MIS problem has several interesting applications, such as portfolio diversification\,\cite{BOGINSKI05} in finance, or broadcast systems (wifi or cellular network) optimization\,\cite{Hale}. If one assigns the value $z_j=+1$ ($z_j=0$) if the vertex $j$ is inside (outside) a given subset, one corresponding objective function is given by 
$C(\bm{z})=-\Delta \sum_j z_j+U\sum_{\langle i,j \rangle}z_i z_j$, in the regime $U\gg \Delta$. Here, the notation $\langle i,j \rangle$ indicates that the two vertices with labels $i$ and $j$ are connected by an edge. $\Delta>0$ is a linear bias towards large sets while $U$ represents the penalty to the cost function induced by the occupation of two connected vertices. The MIS corresponds to the minimum of $C$. A particular MIS problem is the so-called UD-MIS problem (for Unit Disk-MIS), for which two vertices of a 2D graph are connected by an edge if the distance between them is smaller than a fixed given value. This particular combinatorial problem, which is still NP-complete\,\cite{Clark1990}, is the main purpose of this study. Following Ref.\,\cite{Pichler2018}, we introduce below how one can implement a variational algorithm to solve it with a quantum processor made of neutral atoms.\\ 

For combinatorial optimization problems, a common choice for $H_C$ is the diagonal operator in the computational basis defined by $H_C|\bm{z}\rangle = C(\bm{z})|\bm{z}\rangle$, while the mixer Hamiltonian $H_M$ induces transitions between computational basis states\,\cite{Farhi2014,Farhi20142,Hadfield2017}. The Hamiltonian $H_C$ corresponding to the objective function $C$ of a UD-MIS problem in 2D can be realized naturally in ensembles of Rydberg atoms\,\cite{Browaeys2016,Labuhn2016,Pichler2018}. More specifically, for each atom, one take for the state $|1\rangle$ a selected Rydberg state and for the $|0\rangle$ state an atomic ground state. Driving the $|0\rangle\leftrightarrow|1\rangle$ transition with a laser system, the dynamics of the ensemble of atoms is governed in the frame rotating at the laser frequency by the Hamiltonian 
\begin{equation}
    H(t)=\hbar \Omega(t) \sum_j \sigma_j^x-\hbar \Delta(t)\sum_j  n_j+\sum_{ji}\frac{C_6}{r_{ij}^6} n_i n_j.
    \label{ising}
\end{equation}
Here, $n_j|z_j\rangle =z_j|z_j\rangle$ counts the presence of atom $j$ in the Rydberg state and $\sigma^x=|0\rangle \langle1|+|1\rangle \langle 0|$ induces transitions between the states $|0\rangle$ and $|1\rangle$. $\Omega$ is the Rabi frequency of the laser system and $\Delta \geq 0$ is the corresponding detuning between the laser frequency and the bare atomic frequency. The last term in Eq. (\ref{ising}) corresponds to Van der Waals interactions between extended Rydberg states, with $r_{ij}$ denoting the distance between atoms $i$ and $j$. Physically, adding an excitation to the system reduces the total energy of the system by $\hbar \Delta$, but two nearby excitations separated by $r_{ij}$ interact and increase the total energy by $\frac{C_6}{r_{ij}^6}$.  If one approximates the Rydberg-Rydberg interaction by a hard-sphere infinite interaction with characteristic Rydberg blockade radius $r_b\sim \left[C_6/\sqrt{(2\hbar\Omega)^2+(\hbar\Delta)^2}\right]^{1/6}$, the dynamics induced by the Hamiltonian $H$ is restricted to independent sets of the UD-graph naturally defined by the atoms, as illustrated in Fig. \,\ref{UD} [see Appendix \ref{appendix_Rydberg} for details]. One can realize the objective Hamiltonian $H \sim H_C$ by taking $\Omega=0$, $\Delta=\Delta_0 \neq 0$ and the mixer Hamiltonian $H \sim H_M$ by taking $\Omega=\Omega_0\neq 0$,   $\Delta=0$. In the following, we will choose $\Omega_0=\Delta_0$ for simplicity as it does not qualitatively affect the results presented here.\\

The performance of the algorithm can be estimated by the approximation ratio \begin{align}r=\langle C\rangle_f/C_{opt}\end{align}
where $\langle C\rangle_f$ is here the estimate of the objective function at the end of the optimization procedure and $C_{opt}$ the optimal possible value, computed in a separate classical procedure. For randomly placed atoms in a square box of size $L\times L$, the average number of atoms per Rydberg blockade is roughly given by $\nu=N r^2_b/L^2$. When $\nu \ll 1$ (below the percolation threshold\,\cite{Mertens2012}), the UD-graph defined by the atoms can be partitioned in many subgraphs and finding the MIS is easy, as it amounts to summing up the contributions of all the smaller subgraphs. In that regime, there exist efficient classical algorithms that can find the MIS in polynomial time\,\cite{LI2017}. In the opposite regime, $\nu \gg 1$, all the vertices are connected and we trivially have $MIS=1$. In the intermediate regime, the task of finding the MIS is the hardest\,\cite{Pichler2018}. In the following, we will mainly work at these intermediate values of the density.\\

\section{Robustness to spontaneous emission}
\label{section_robustness}

The implementation of the quantum dynamics described in the previous sections on a real quantum processor will suffer from 
imperfections and errors. In the case of a quantum processor made of neutral atoms controlled by laser fields, these errors arise from the combination of several physical effects\,\cite{deLeseleuc2017}, such as Doppler effect, laser phase noise, or spontaneous emission. As illustrated in Appendix \ref{appendix_dissipation}, these processes can be described by an effective master equation describing the density matrix $\rho$ of the system,
\begin{align}
   \partial_t \rho=&-\frac{i}{\hbar}[H(t),\rho]-\frac{\Gamma}{2}\sum_j\left(\sigma_j^+\sigma_j^-\rho+\rho\sigma_j^+\sigma_j^- -2\sigma_j^- \rho \sigma_j^+\right)\notag\\
   &-\frac{\gamma}{2}\sum_j\left(n_j \rho+\rho n_j  - 2 n_j \rho n_j \right).
   \label{master_eq}
\end{align}
Here, $\Gamma$ and $\gamma$ are effective phenomenological spontaneous emission and dephasing rates, $\sigma^+=|1\rangle \langle 0|$ and $\sigma^-=(\sigma^+)^{\dagger}$. In the following, we will compute the dynamics of the system either via the evolution of the wave function under stochastic quantum jumps\,\cite{Daley2014} using an average over a large number of trajectories, or using direct time-evolution of the master equation for the density matrix. For simplicity, we implement here the algorithm including only 3 variational parameters, $(t_1,\tau_1,t_2)$ and always use the same classical optimizer, the Nelder-Mead simplex method\,\cite{Nelder1965} with standard optimization parameters.\\

We find numerically that the variational algorithm under study displays an effect of self-mitigation of errors, in the sense that performances in the presence of noise are better than what we would naively expect. More specifically, we find that the algorithm sometimes spontaneously deviates from the parameters that would be obtained in the absence of dissipation, if the corresponding time-evolution is strongly affected by dissipation. In those cases, the algorithm may elect a new set of parameters (that would be suboptimal in the absence of noise), for which the preparation suffers less from dissipation, resulting in a better final performance. \\

 \begin{figure}[t!]
\center
\includegraphics[scale=0.5]{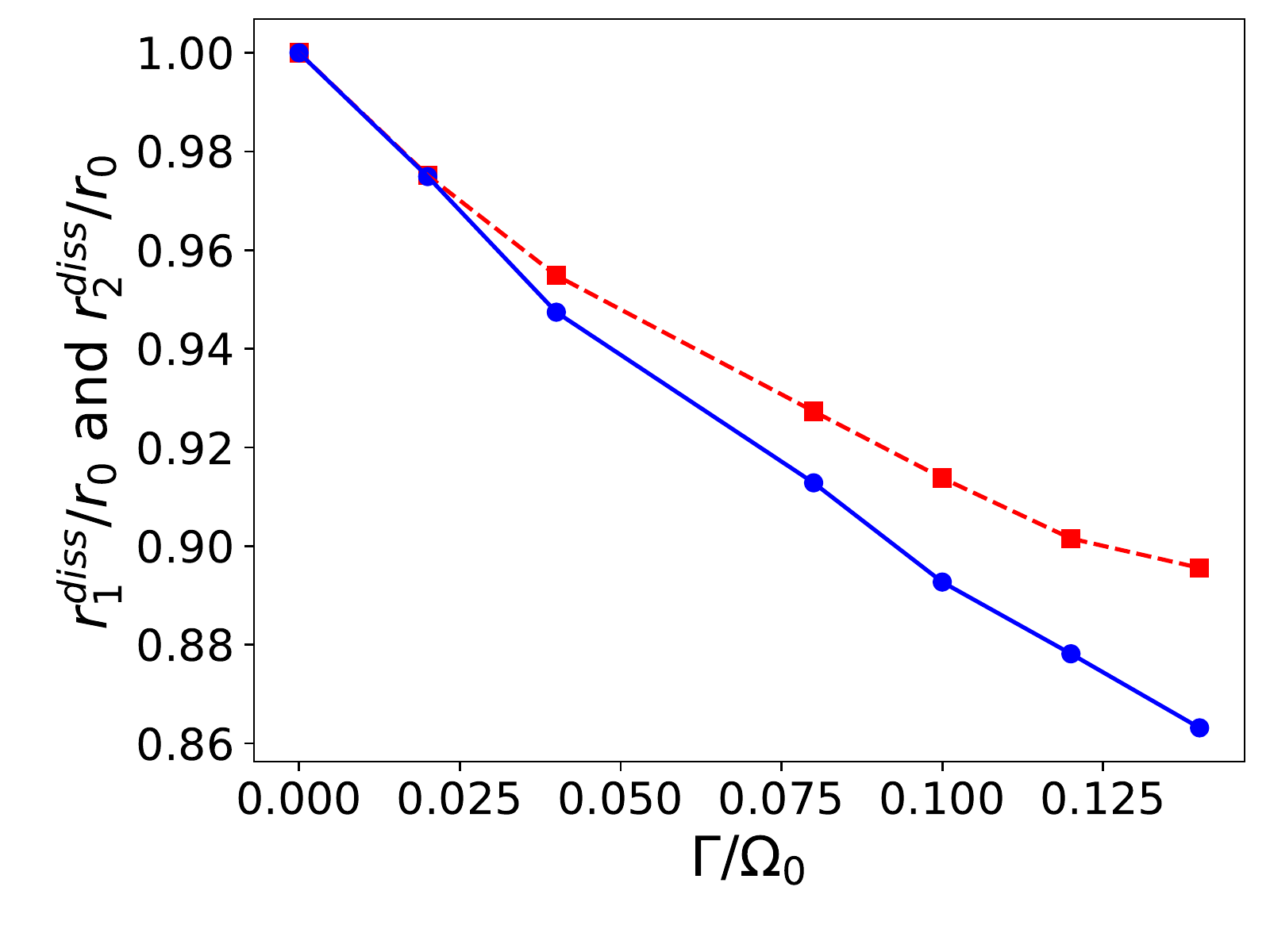}
\caption{Evolution of $r_1^{diss}/r_0$ (blue dots anf full curve) and $r_2^{diss}/r_0$ (red squares and dashed curve) with respect to the spontaneous emission rate $\Gamma/\Omega_0$. All the results are averaged over 100 random graphs of $N=14$ atoms with density $\nu=2.6$. The optimization procedure is done with a small number of variational parameters for simplicity, and involves 5 randomly generated initial conditions, where each individual parameters is sampled uniformly in $[0,\pi/(\hbar\Omega_0)[$. We use a direct time-evolution of the density matrix with Eq. (\ref{master_eq}). }
\label{performances_spontaneous_emission}
\end{figure}

This behavior can be revealed by the following numerical experiment. We randomly generate multiple graph instances with density $\nu=2.6$. For each of these instances, we run the variational algorithm without dissipation, with 5 randomly chosen starting points for the initial parameters. This gives us a particular value for the approximation ratio $r_0$ and optimal parameters $\bm{t}^{0}$ and $\bm{\tau}^{0}$. Then, we apply a time-evolution in the presence of spontaneous emission with the parameters $\bm{t}^{0}$ and $\bm{\tau}^{0}$ and obtain a dissipative approximation ratio $r_1^{diss}<r_0$. Alternatively, we run the variational algorithm with spontaneous emission, i.e. computing the time-evolution with Eq.\,(\ref{master_eq}), with the same 5 starting points as in the non-dissipative case, and obtain another dissipative approximation ratio $r_2^{diss}<r_0$. We plot in Fig.\,\ref{performances_spontaneous_emission} the values of the ratios $r_1^{diss}/r_0$ (blue dots) and  $r_2^{diss}/r_0$ (red squares), averaged over multiple graph instances, as a function of the spontaneous emission rate $\Gamma$ ($\gamma=0$ here). As shown on the figure, one obtains quantitatively better results when the algorithm explores the parameter space freely. \\


 \begin{figure}[t!]
\center
\includegraphics[scale=0.55]{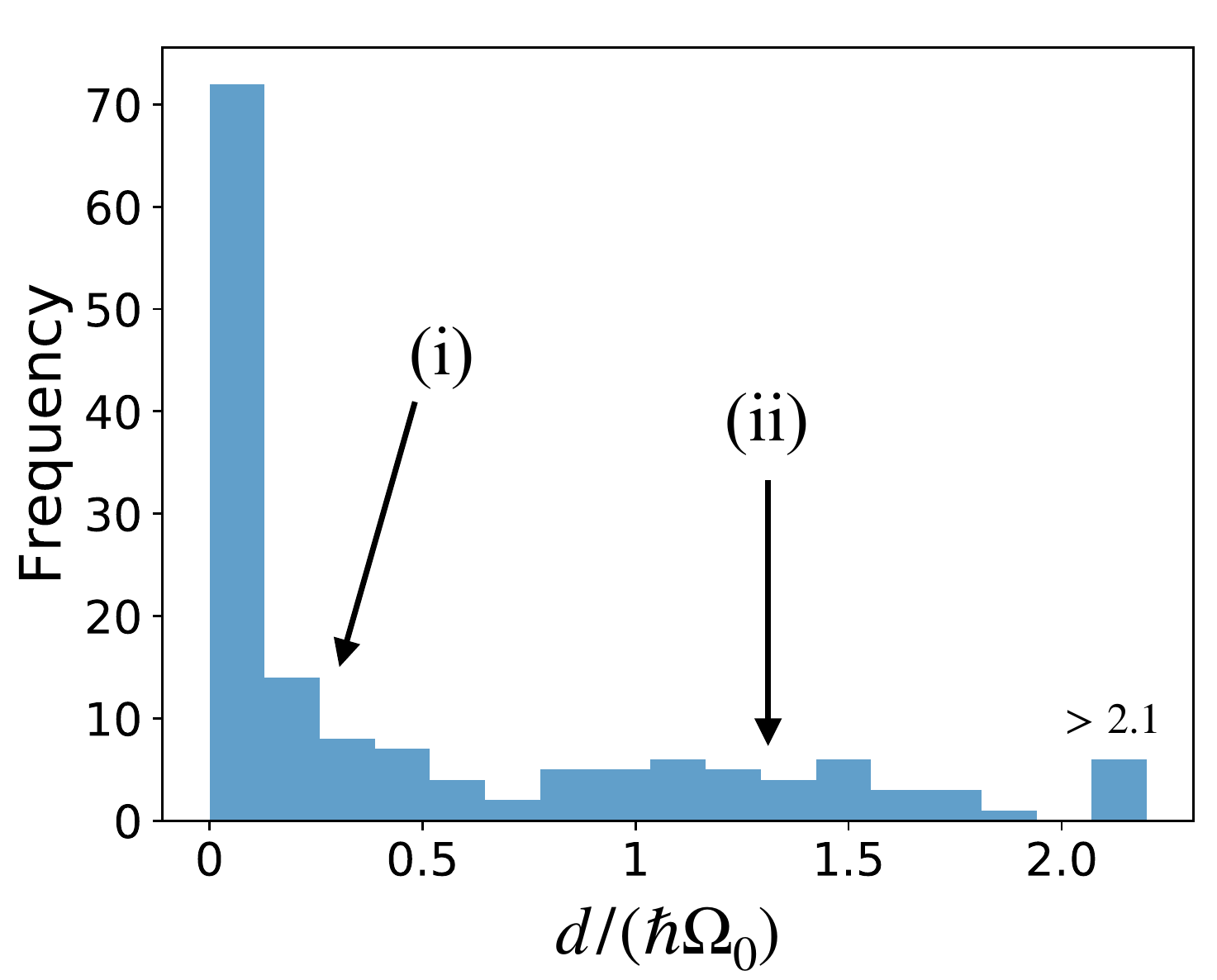}
\caption{Histogram $d/(\hbar\Omega_0)$ for 154 randomly generated graph instances with $N=18$ atoms and a density $\nu=2.6$, starting from the same 15 initial parameters for both the noisy and the noise-free optimization procedures. The histogram displays a large tail (region (ii)), for which the optimal parameters in the presence of spontaneous emission differ quantitatively from the ones obtained with an ideal time-evolution. We use an average over 2000 wavefunction trajectories for the evaluation of the dynamics of the system with Eq. (\ref{master_eq}).}
\label{Histogram}
\end{figure}

To confirm that the dissipative optimization procedure sometimes results in a quantitative change of the optimal parameters $\bm{\tau}$ and $\bm{t}$, we conduct a separate numerical experiment at a fixed spontaneous emission rate $\Gamma/\Omega_0=0.1$. We randomly generate multiple graph instances with $N=18$ atoms and $\nu=2.6$, and compute in each case the Euclidean distance $d=[\sum_{j} (t_j^{diss}-t_j^0)^2+\sum_{j}(\tau_j^{diss}-\tau_j^0)^2]^{1/2}$ between the parameters $\bm{\tau}^{diss}$ and $\bm{t}^{diss}$ obtained at the end of the optimization procedure with dissipation and the parameters $\bm{\tau}^{0}$ and $\bm{t}^{0}$ obtained in the absence of dissipation. In all cases, we repeat the optimization procedure 15 times, each time with random initial parameters which are identical for both the noisy and the noise-free optimization procedures. We show in Fig.\,\ref{Histogram} the histogram of the results that are obtained. In most cases, we obtain a rather small distance $d$ close to zero, as illustrated by the points in the region (i). For these instances, the optimal parameters of the dissipative optimization remain close to those with the dissipation-free optimization. In addition, the histogram displays a rather large tail at larger values $d\sim (\hbar \Omega_0)^{-1}$, corresponding to the region (ii). For those instances, the dissipative optimization procedure converges to optimal parameters that differ quantitatively from the ones obtained without dissipation. In those cases, spontaneous emission modifies the optimization landscape in such a way that a novel optimum is reached.\\

 \begin{figure}[t!]
\center
\includegraphics[scale=0.5]{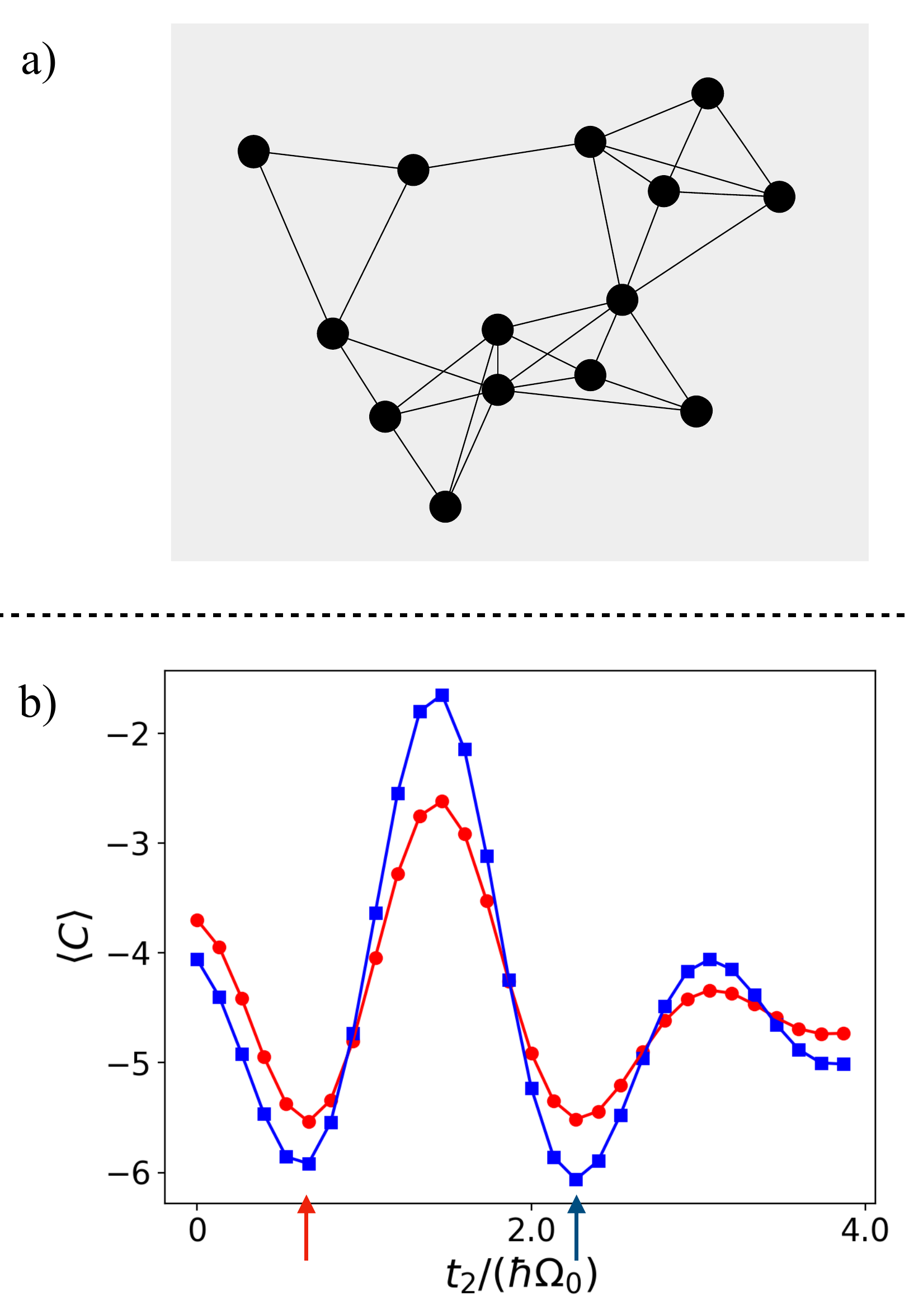}
\caption{We show on panel a) one particular UD graph (the positions of the vertices have been modified for more visibility, but the links remain unchanged). The panel b) shows the evolution objective function with respect to the parameter $t_2$ for $\Gamma=0$ (blue squares) and $\Gamma/\Omega_0=0.1$ (red dots). We have taken $t_1/(\hbar\Omega_0)=1.5$ and $\tau_1/(\hbar\Omega_0)=1.$. The noisy time-evolution is averaged over 10$^3$ quantum trajectories.   }
\label{Example_graph}
\end{figure}

Typically, shorter time-evolution preparations are less affected by spontaneous emission, and may therefore become more favorable in comparison with longer ones. This is exemplified in Fig \ref{Example_graph}, where we show a particular graph example for which one can directly witness the effect of dissipation. The graph under consideration is displayed on panel a), and was generated randomly with density $\nu=2.6$. For illustration purposes, we keep the first parameters $t_1/(\hbar \Omega_0)=1.5$ and $\tau_1/(\hbar \Omega_0)=1.0$ fixed, and we study the evolution of the objective function $\langle C \rangle$ to be minimized with respect to $t_2$ only. We show on the panel b) of Fig. \,\ref{Example_graph} the evolution of $\langle C \rangle$ with respect to $t_2$, with (red dots) and without (blue squares) spontaneous emission. In the absence of spontaneous emission, the optimal parameter $t_2$ would be around $t_2/(\hbar\Omega_0)\sim 2.2$ (see blue arrow), whereas one obtains a far smaller value in the dissipative case (see red arrow).\\

It is important to note that this effect might lead to a failure of standard extrapolation techniques to zero-dissipation. In particular, if one has only access to the minima of the objective function when dissipation is large, the extrapolation will not converge to the global minimum of the dissipationless curve at $t_2/(\hbar\Omega_0)\sim 2.2$, but rather to the one at $t_2/(\hbar\Omega_0)\sim 0.6$.  \\

Note that the current experimental capabilities\,\cite{Sylvainthese} allow us to work with quite small values of $\Gamma/(\hbar\Omega_0)< 10^{-2}$, for which the effect described here is rather small (see Fig. \ref{performances_spontaneous_emission}). We belive that this effect may however play a role when scaling up to larger atom numbers.\\

We have seen above that the algorithm under study was relatively robust to spontaneous emission and dephasing processes. The general nature of this effect should also be applicable to other variational algorithms pursuing approximate resolution of problems. It therefore motivates their study for implementation on near-term quantum processors. In the following Section, we present some improvements of the procedure.

\section{Improvements of the optimization procedure}

\begin{figure}[t!]
\center
\includegraphics[scale=0.45]{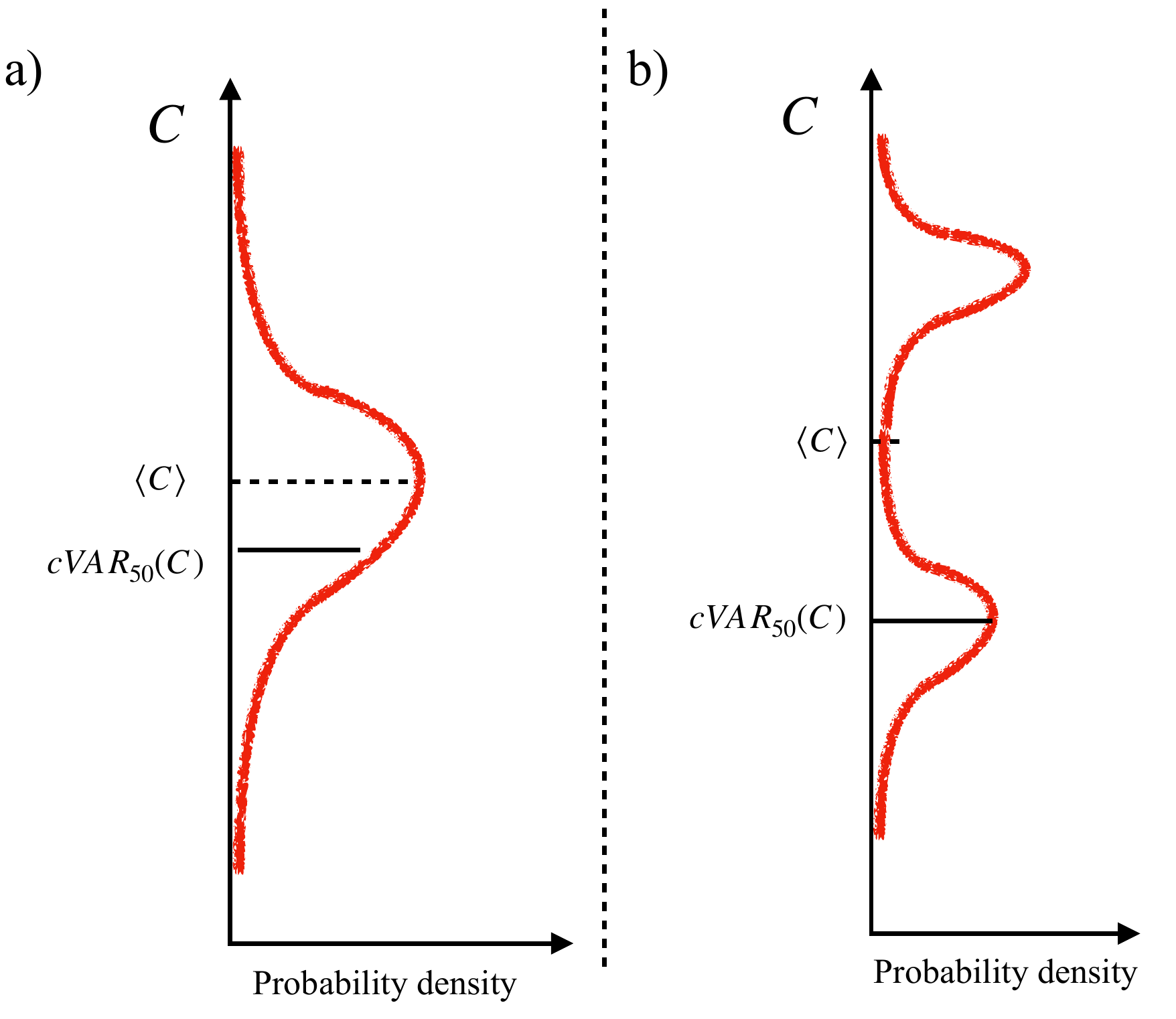}
\caption{Sketches of the probability density associated with the measurement of two  distinct states $\rho_a$ (a) and $\rho_b$ (b) in the computational basis. The preparation of state $\rho_b$ would be favored with respect to the preparation of $\rho_a$ when using the cVAR objective function. }
\label{cVAR}
\end{figure}

It has recently been suggested to use other observables instead of the mean energy during the minimization procedure\,\cite{Barkoutsos2019}. This proposal comes from the fact that we are solely interested in the properties of the minimal energy state, whereas the quantum processor might only be able to prepare a superposition state in the measurement basis, therefore leading to poor average performances. One obvious way to focus on the ground state properties would be to choose for the objective function the minimum observed energy over a set of measurements. However, this approach would generate an irregular and uneven objective function that would be difficult to deal with in the classical optimization loop. To render the objective function more regular, while still favoring the best measured outcomes rather than the mean value, one can work with the conditional value at risk (cVAR$_m$, or expected shortfall) corresponding to the mean value of the energy in the $m\%$ most favorable cases. This quantity, which is widely used in risk-analysis in the financial sector\,\cite{cvar-finance}, permits then to focus on the tail of the energy distribution. Using the cVAR$_m$ objective function allows us to reach greatly enhanced performances, by tailoring the objective function to the specific task of having a relatively large overlap with low energy states. This is illustrated in Fig.\,\ref{cVAR}, where we sketch on panels a) and b) the two measurement probability distributions associated with two distinct quantum states $\rho_a$ and $\rho_b$. With an optimisation procedure on the mean value of the energy $\langle C\rangle$, the state $\rho_a$ would be preferred. On the other hand, using cVAR$_{50}$ as the objective function would result in the preparation of $\rho_b$, that has a larger overlap with low energy states. 

\begin{figure}[t!]
\center
\includegraphics[scale=0.5]{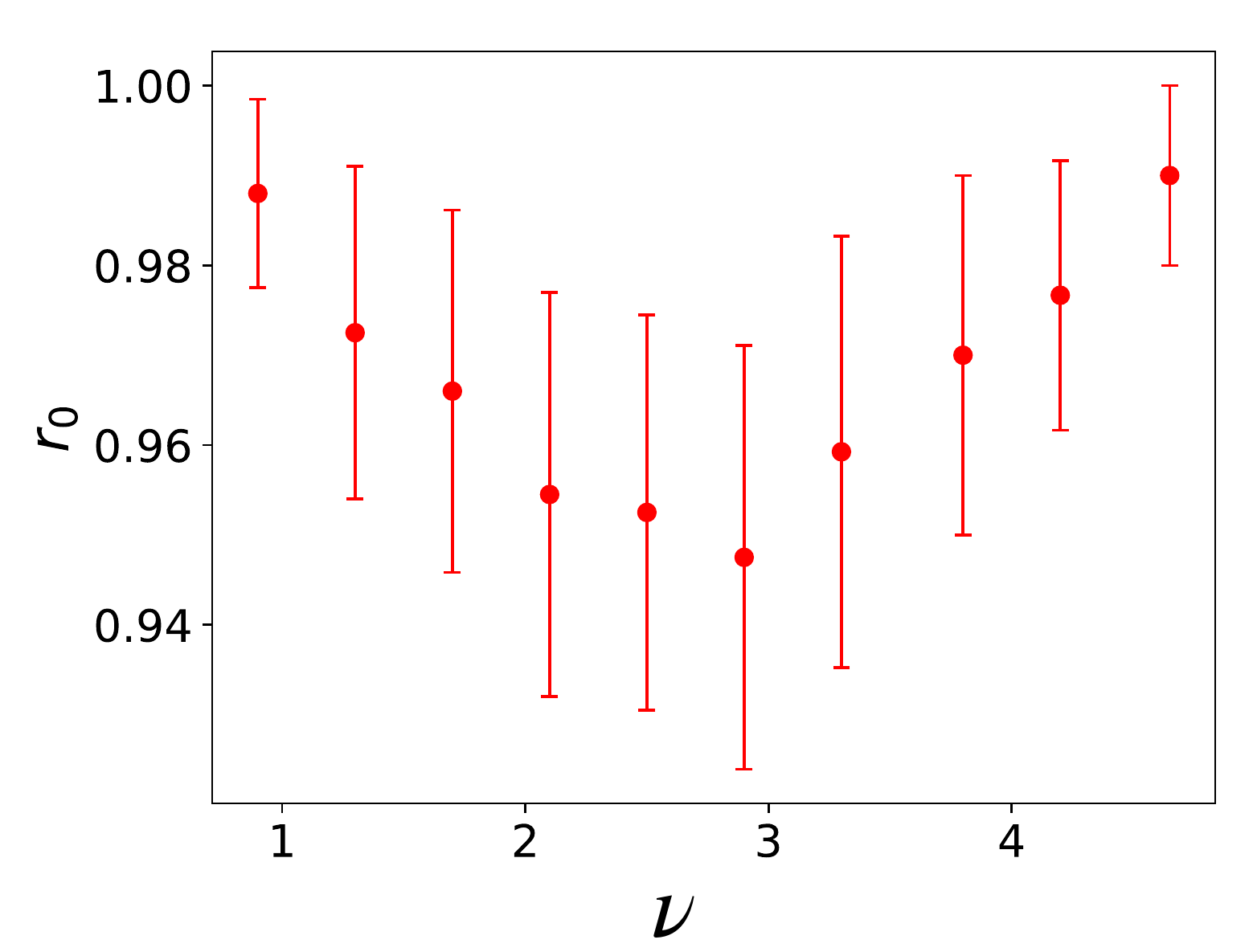}
\caption{We show the evolution of the approximation ratio $r_0$ (and its standard deviation) as a function of the density $\nu$. We have a three variational parameters and $N=18$ atoms, and each data point is averaged over 50 graph realizations, starting from the initial state with parameters initialized at $\pi/(\hbar \Omega_0)$.}
\label{Performances}
\end{figure}

 With this choice, the algorithm displays very good performances $r_0 \geq 0.9$ already when initialized from a unique initial condition, as we illustrate in Fig.\,\ref{Performances}, where we plot the approximation ratio $r_0$ with respect to the density $\nu$, for $N=18$ atoms, $m=20\%$. Fig.\,\ref{Performances} also illustrates how the performance of the algorithm varies with the density. At low densities, the graph instances can be partitioned in multiple subgraphs, for which the task of finding the MIS is computationally facilitated. At large densities, graph instances are very connected and the number of independent sets is greatly reduced, thereby leading to good search performances of the MIS. As expected from these arguments, the resulting plot in Fig.\,\ref{Performances} exhibits minimal performances at intermediate densities $\nu\sim 3$. For each density, the bars show the standard deviation obtained over the 50 random graph realizations. The minimal average performance is also associated with a rather large standard deviation, illustrating the occurrence of a few hard graph instances associated with relatively bad performances of the algorithm.\\

Parameter initialization is another aspect that could be improved upon in order to enhance the performances of the algorithm. It has notably being proposed to use meta-learning procedures\,\cite{Verdon2019} for the algorithm to learn how to best initialize the variational parameters. Combinations of classical pre-processing with heuristic methods for parameter initialization and optimization might then allow to maintain good levels of performances when increasing the number of atoms, while keeping the number of variational parameters and total runtime low.

\section{Conclusion}

In this paper, we have demonstrated an effect of self-mitigation of noise in an approximate variational quantum algorithm applied the the MIS problem, implemented with arrays of neutral atoms. This mechanism originates from the non-uniform effect of incoherent errors on the optimization landscape. Due to this interpretation, this generic mechanism should be readily generalizable to other variational approximate procedures. As such, the presence of decoherence in NISQ quantum computers does not represent a daunting obstacle towards a possible quantum advantage with variational algorithms. In addition, we illustrated that overall performances can be further improved by changing the objective function to be optimized for, motivating the development of heuristic methods to maximize the performances of variational NISQ algorithms.

\section{Acknowledgements}
This work benefited from discussions with L. Beguin, A. Browaeys, C. Jurczak, T. Lahaye, P. Remy, G.-O. Reymond, and A. Signoles. This work was granted access to the HPC resources of CINES under the
allocation 2019- A0070911024 made by GENCI.

\appendix

\section{Platform of Rydberg atoms\label{appendix_Rydberg}}

\begin{figure}[t!]
\center
\includegraphics[scale=0.31]{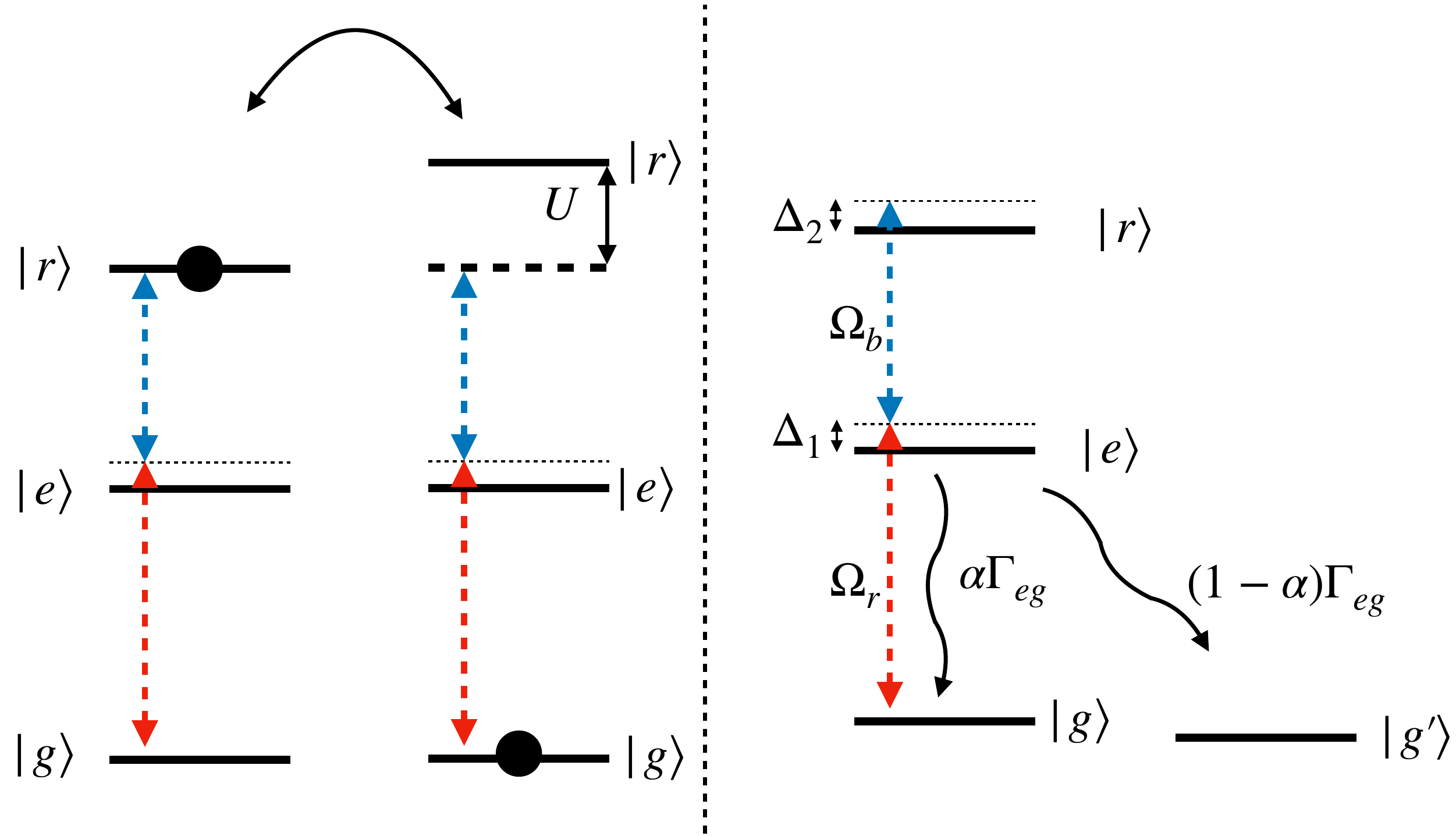}
\caption{Left panel : Illustration of the Rydberg blockade effect. The presence of the first atom in the Rydberg state induces a strong energy shift for the second atom at short distance smaller than $r_b$, thereby putting the Rydberg level strongly off-resonance with the laser system.}
\label{Blockade}
\end{figure}

Recent experimental developments allowed for the realization of large two-dimensional\,\cite{Barredo2016} and three-dimensional\,\cite{Barredo18} arbitrarily-shaped arrays of single Rydberg atoms separated by a few micrometers. One can encode a two-level system between the ground state
and a selected Rydberg level. The van der Waals interaction
between two atoms in Rydberg states gives rise to the Ising-like Hamiltonian in Eq.\,(\ref{ising}) of the main text. In practice, we couple the ground and Rydberg state through an intermediary state $|e\rangle$ with a two-tone laser system, as illustrated in Fig.\,\ref{Blockade}. Strong Van der Waals interaction gives rise to the phenomenon of Rydberg blockade, where two atoms separated by a distance smaller than the typical interaction length $r_b$ of the interaction cannot be excited in the Rydberg state at the same time. This phenomenon is illustrated on the left panel of Fig.\,\ref{Blockade}.

\section{Modelling of dissipation\label{appendix_dissipation}}

We present in this Appendix how we incorporate in the numerical simulations the effects of dissipation and imperfections.

\subsection{Spontaneous emission}

The minimal realistic modelling of the system\,\cite{Sylvainthese} involves a 4-level structure, as depicted on the right panel of Fig.\,\ref{Blockade}. In that picture, $|g\rangle$ and $|r\rangle$ are coupled via a two-photon transition, with corresponding Rabi frequencies $\Omega_r$ and $\Omega_b$ and single and two-photon detunings $\Delta_1$ and $\Delta_2$ (see Fig.\,\ref{Blockade}, left panel). The intermediate state $|e\rangle$ has a finite lifetime $1/\Gamma_{eg}$, and spontaneous emission can bring back the atom either to the ground state $|g\rangle$ with probability $\alpha=1/3$ or to another uncoupled state $|g'\rangle$ with probability $2/3$. The Rydberg state is considered to have an infinite lifetime. \\

In terms of numerical complexity, it is highly desirable to have an effective description of the dynamics in terms of an effective 2-level system. For $\alpha=1$ (i.e. a standard 3-level system), one can integrate out the intermediate level in the regime $\Delta_1 \gg \Omega_r,\Omega_b$. After this elimination, one finds that $|g\rangle$ and $|r\rangle$ are effectively coupled by an effective field with Rabi frequency $\Omega=\Omega_r \Omega_b/(2\Delta_1)$ and detuning $\Delta=\Delta_2+(\Omega_r^2-\Omega_b^2)/(4\Delta_1^2)$. After this procedure, the $|r\rangle$ level acquires a finite decay rate $\Gamma=\Gamma_{eg} (\Omega_r^2+\Omega_b^2)/(4\Delta_1^2)$. Throughout the paper, we use this simplified description with $\alpha=1$, as details of the atomic structure are not directly relevant to the physical effects that we describe.

\subsection{Dephasing}

The experimental system features several other sources of imperfections than spontaneous emission. First, the atoms have different non-zero velocities that result in a spread of the effective laser frequency effectively seen by the atoms, through the Doppler effect. Moreover, the excitation lasers that we use are not purely monochromatic but instead display some phase noise. One can take these effects into account with a phenomenological local dephasing model with a rate $\gamma$, as presented in Ref.\,\cite{Lienhard2018}. Comparable results are obtained in this case.




\bibliography{refs3.bib}
\end{document}